\documentstyle[psfig]{mn}
\input{psfig.sty}

\newif\ifAMStwofonts

\title[Quiescent times in $\gamma$-ray bursts]
        {Quiescent times in $\gamma$-ray bursts: II. Dormant periods
        in the central engine?}
\author[Ramirez-Ruiz, Merloni \& Rees]
        {Enrico Ramirez-Ruiz, Andrea Merloni and  Martin J. Rees
\\Institute of Astronomy, Madingley Road, Cambridge, CB3 0HA}

\date{}

\begin{document}

\maketitle

\label{firstpage}

\begin{abstract}
Within the framework of the internal-external shocks model for
$\gamma$-ray bursts, we study the various mechanisms
that can give rise to  
quiescent times in the observed $\gamma$-ray light-curves.
In particular, we look for the signatures that
can provide us with evidence as to whether
or not the central engine goes dormant for a period of time comparable to the
duration of the gaps.           
We show that the properties of the prompt $\gamma$-ray and X-ray
emission can in principle determine whether the quiescent episodes are due 
to a modulated relativistic wind or a switching off of the central engine. 
We suggest that detailed observations of the prompt 
afterglow emission from the reverse shock will strongly constrain the possible
mechanisms for the production of quiescent times in $\gamma$-ray bursts.
\end{abstract}


\section{Introduction}
The origin of $\gamma$-ray bursts (GRBs) has been one of the great unsolved
mysteries in high-energy astrophysics for almost 30 years. The recent
discovery of fading sources at X-ray \cite{costa97} and optical \cite{vp97}
wavelengths has established that GRBs lie at cosmological distances,
making them the most luminous events known in the Universe. During their brief
duration, their photon luminosities exceed by many
orders of magnitude the most extreme output from any active galactic
nucleus. However, the total energy budget needed to produce a GRB 
is not beyond the scope of some other phenomena encountered in astrophysics.

The two most popular models to
explain GRBs are the coalescence of two compact objects
such as black holes or neutron stars \cite{lat76}, or the cataclysmic
collapse of a massive star in a very energetic supernova-like explosion
\cite{mc99,pac98}. The formation of a black hole with a debris torus around it
is a common ingredient of both these scenarios. The binding energy of
the orbiting debris and the spin energy of the black hole are the two
main reservoirs available, the extractable energy being up to
$10^{54}$ ergs \cite{Rees99}.\\

A key issue that has remained largely unexplained is what
determines the characteristic durations of the bursts (typically
between $10^{-2}$ and $10^{3}$ seconds). While bursts lasting
hundredths of a second could be derived from a very short, impulsive energy
input, this is generally unable to account for the complicated
temporal structure found in a large fraction of the $\gamma$-ray
burst light-curves \cite{piran97,fen99,err99}. This is suggestive of a
central engine that releases energy, in the form of a wind or multiple
shells, over a period of time commensurate with the observed duration of a GRB
\cite{Rees94}. 

The lack of apparent photon-photon attenuation of high energy
photons implies substantial bulk relativistic motion. The relativistic
shells must have a Lorentz factor, $\Gamma =
(1-\beta^{2})^{-1/2}$, of the order of 10$^{2}$ - 10$^{3}$ (note that even 
if -- as is likely -- the outflow is beamed the spherical shell model is 
applicable provided the beam is wider than an angle $\sim \Gamma^{-1}$).
 The observed afterglow emission in this scenario is
produced when the expanding shells slow down as a result of the
interaction with the surrounding medium.
 
This internal-external fireball model
\cite{mes97,piran97} requires a complicated central engine which
persists (though unsteadily) for much  longer than the typical
dynamical time scale of  a stellar mass compact
object, which is of the order of  milliseconds.

Observationally, there is a bimodal distribution of burst 
durations (Kouveliotou et al. 1993). 
It is a plausible conjecture that the short ($\la 1$ s)
bursts are triggered by coalescing compact binaries, and the long ones by
a special kind of supernova-like explosion (we note, furthermore, that those 
with detected afterglows are all in the long category; this is
a selection effect caused by the fact that BeppoSAX has a $\sim 5$
second trigger).

In the internal shocks scenario for GRBs, the actual burst 
temporal profile is the
outcome of the complex dynamics of the ejecta, which have usually 
been treated as concentric shells moving at different speeds. 
Although a
range of Lorentz factors seems likely, it is not obvious
whether most of the energy (or the mass) would be concentrated towards the
high or low end of the $\Gamma$ distribution. In this early phase, the
time-scale of the burst and its   
overall structure follows, to a large extent, the temporal behaviour of
the source \cite{koba97}.
In contrast, the subsequent afterglow
emerges from the shocked regions of the external medium where the relativistic 
flow is slowed down; therefore the inner engine cannot be seen
directly in the afterglow. The external medium may be typical
interstellar matter (ISM; $\rho_{\rm ext} \sim 1$ cm$^{-3}$), 
but it would be much denser if a massive star 
underwent rapid mass-loss before the burst was triggered 
($\rho_{\rm ext} \ga 10^4$ cm$^{-3}$, see Ramirez-Ruiz et al. 2001).  
It is thus of great importance to obtain as much information as possible on
the nature of the early relativistic outflow, as this would provide us
with some of the best 
clues to the nature of GRB progenitors. 

Long GRBs, which are very
complicated in the time domain, may also show multiple episodes of emission,
separated by background intervals  or
quiescent times of variable duration. In earlier
discussions (Fenimore \& Ramirez-Ruiz 2000, Ramirez-Ruiz \& Merloni
2001, hereafter Paper I), the  
presence of quiescent times has been regarded  as an indication of a
turning-off of the central site for a period of time. 
However,  if gamma-ray bursts are produced by internal shocks in
relativistic winds,  it is possible as an alternative
to attribute the 
quiescent times in  the $\gamma$-ray light-curves  to a
complicated  modulation  of the ejecta velocities in the relativistic
outflow.
 
In this paper we explore the feasibility of different
internal-external  shock models to produce episodes of
quiescence in a $\gamma$-ray burst. We calculate the expected
light-curves of the early multi-wavelength emission. We show that
light-curves of the prompt afterglow in the optical, X-ray and
$\gamma$-ray bands could  provide us with strong evidence
as to whether or not the central engine goes dormant for a period of
time comparable to the duration of the gap.

\section{Quiescent Times in Internal Shocks}

\subsection{Model outline}

We simulate GRB light-curves by adding pulses radiated in a series of
internal shocks that occur in a transient, unsteady relativistic
wind. Several authors have modelled this process by randomly selecting
the initial conditions at the central site
\cite{daigne98,koba97,SPM00}. Here we model the wind dynamics and the 
emission processes as in Fenimore \& Ramirez-Ruiz (2000), but we include
the effect of the 
photon diffusion through the colliding shells and the wind on the
pulse duration \cite{SPM00}. 

As described in Fenimore \& Ramirez-Ruiz (2000), the wind is discretized in a
sequence of N shells that are ejected over a period $T_{\rm dur}$ from
the central source, with a range of initial thicknesses ($\Delta_i$). We
randomly select $t_{0i+1}-t_{0i}$ from a Poisson distribution based
on the rate 
of peak occurrence. Thus, we specify the rate of explosions at the
central site (${N \over T_{\rm dur}}$) in such a way that the actual number of
peaks is random. We generate about 1.4 shells per second\footnote{The
substantial overlap of the temporal structures in the burst have made
the study of individual pulses somewhat difficult. An excellent
analysis has been provided by Norris et al. (1996), who examined the
temporal structure of bright GRBs by fitting time histories with
pulses. They found that the distribution of intervals between pulses
exhibits a broad maximum near 0.8 s.}.   
We set the maximum thickness to be $\Delta_i \sim 0.3$ lt-s. 
The peak energy can be
estimated from the bursts with measured redshifts. GRB970508 had a peak
luminosity  $L \sim 3 \times 10^{51}$ erg s$^{-1}$. Other
GRBs have been found at extreme redshifts \cite{kul99}, implying $L \sim 2
\times 10^{53}$ erg s$^{-1}$.   
In our simulations the shell peak energy (isotropic equivalent) 
$E_i/4\pi$ is drawn from a log-normal
distribution with an average value $\overline{E} = 10^{51}$ erg
s$^{-1}$ and a dispersion $\sigma_E = 10^{1.5}$ erg
s$^{-1}$, thus allowing the occasional ejection of very energetic
shells.

\begin{figure}
\vbox to105mm{\vfil 
\psfig{figure=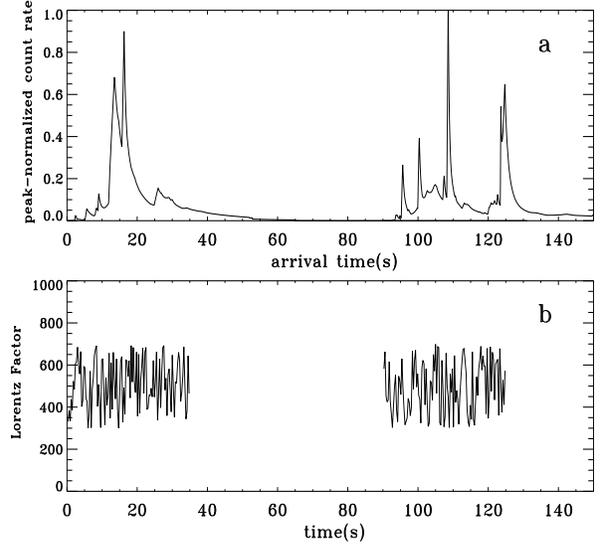,angle=0,height=80mm,width=100mm}
\caption{Simulations of the effects of turning off 
the central source in an internal shock model. Panel (a) shows 
the light-curve at
the detector generated by a central engine which emits shells at the
mean rate of 1.4 per second, with
Lorentz factors chosen randomly between  $3 \times 
10^{2}$ and $9\times 10^{2}$, and turns off between $t=35$ s and
$t=90$ s, as shown in panel (b) ($T_{\rm dur}$=125 s). The external density 
has been fixed to 1 cm$^{-3}$.}
\vfil}
\label{fig1}
\end{figure}

\begin{figure*}
\vbox to145mm{\vfil 
\psfig{figure=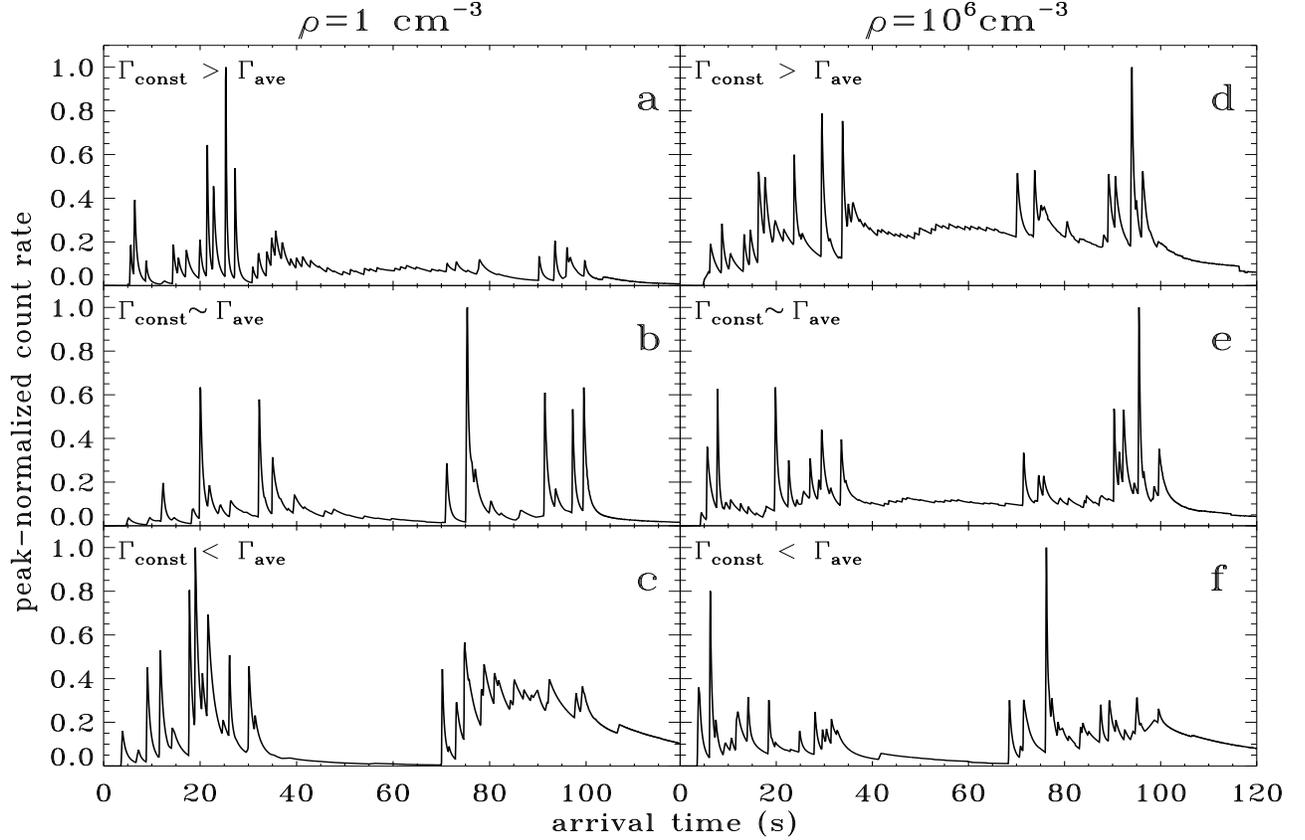,angle=0,height=115mm,width=183mm}
\caption{Consequences of different steady state
scenarios observed in the $\gamma$-ray profiles for a burst triggered in
two different density  environments. The central
engine ejects shells with a constant
Lorentz factor, $\Gamma_{\rm const}$, between $t={1 \over 3}T_{\rm dur}$ and
$t={2 \over 3}T_{\rm dur}$ ($T_{\rm dur}=100$ s). 
The rest of the time the Lorentz factors are
randomly selected between $10^{2}$ and $10^{3}$. The
calculated light-curves for the cases
$\Gamma_{\rm const} > \Gamma_{\rm ave} = 550$ (a and d),
$\Gamma_{\rm const} \sim \Gamma_{\rm ave}$ (b and e) , and
$\Gamma_{\rm const} < \Gamma_{\rm ave}$ (c and f) are shown for
$\rho_0$=1 and $\rho_0$=10$^{6}$ respectively.   
A quiescent time could be observed when the Lorentz factor of the
steady flow is much smaller than 
the average Lorentz factor. The level of the underlying smooth
component produced by collisions with the steady ejecta increases as
the product $\rho_0 \Gamma_{\rm const}^2$ is increased.}
\vfil}
\label{fig2}
\end{figure*}

We calculate the radii where shells collide 
and determine the emission features for each
pulse. If some inner shell moves faster than
an outer one ($\Gamma_i > \Gamma_j$), it will overtake the slower one
at a radius $R_i(t_{ij})=R_j(t_{ij})=R_{\rm c}$. The resulting pulse
reaches the detector at the relative time of arrival,

\begin{figure*}
\vbox to145mm{\vfil 
\psfig{figure=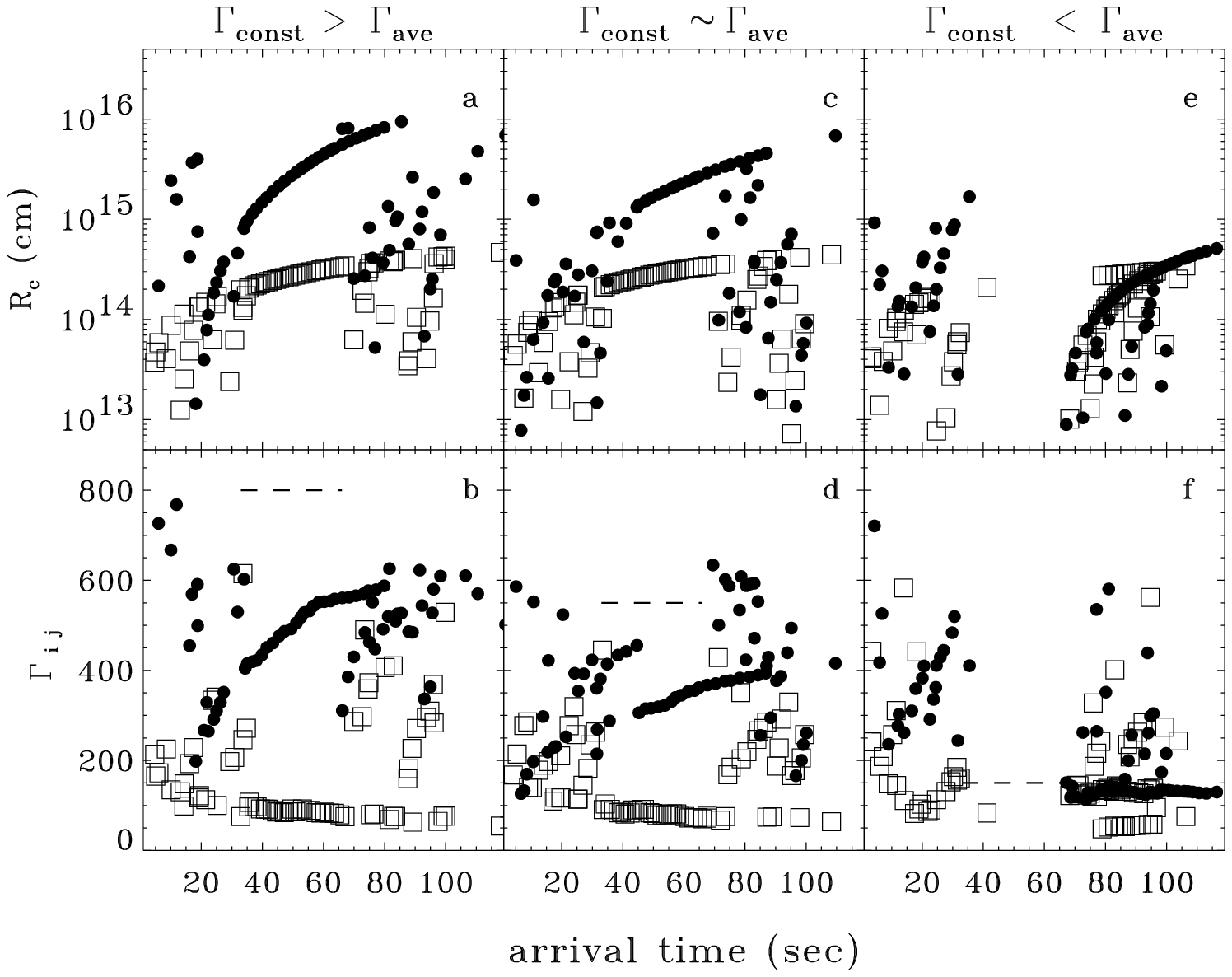,angle=0,height=115mm,width=183mm}
\caption{Collisional parameters of the internal shocks produced 
in the simulations of three  steady outflow cases 
(see text for more details) corresponding to the 
light-curves shown in Fig. 2.
Filled circles correspond to the low density case ($\rho_0=1$) and open
squares to the high density case ($\rho_0=10^6$).
Plotted are the radii $R_c$ where the
collisions take place and the resulting Lorentz factors $\Gamma_{ij}$ as functions of
the relative times of arrival at the detector, $T_{\rm toa}$. 
The radii of collision for a
central engine that ejects, during its steady state, faster shells
($\Gamma_{\rm const} > \Gamma_{\rm ave} = 550$), average speed shells 
($\Gamma_{\rm const} \sim \Gamma_{\rm ave}$), 
and slower shells ($\Gamma_{\rm const} < \Gamma_{\rm ave}$), 
are shown in panels (a), (c) and (e)
respectively. The resulting Lorentz factors as functions of
arrival time  for each of these scenarios are shown in  panels (b),
(d) and (f), respectively. The dashed lines in panels (b), (d) and (f) mark 
the value of $\Gamma_{\rm const}$ in the three cases.}
\vfil}
\label{fig3}
\end{figure*}

\begin{equation}
T_{\rm toa}=t_{ij}-{R_{\rm c} \over c}=t_{0i}+ {\Gamma_j^2 \over \Gamma_i^2 -
\Gamma_j^2}\Delta t_{0ij},
\label{toa}
\end{equation}
where $\Delta t_{0ij}=t_{0i}-t_{0j}$ is the difference
between the times at which the inner engine generates the { \it i}-th
($t_{0i}$) and the { \it j}-th shell ($t_{0j}$). The collision
radius, $R_c$, is roughly  
\begin{equation}
R_c \sim c\Gamma^2\Delta T,
\label{radius}
\end{equation}
where $\Delta T$ is the typical time variation observed in a GRB
\cite{Rees94}. For typical values such as  $\Delta T=0.1$ to $1.0$ s,
$R_c$ is about $10^{14}$cm. As the fast later shells move outward
they begin to interact with the external medium and decelerate. The
deceleration is expected to occur at
\begin{equation}
R_{\rm dec}=10^{16.7}({ E_{52} \over \rho_0 \theta^2})^{1/3}\Gamma_2^{-2/3} 
{\rm cm},
\label{rdec}
\end{equation}
where $E_0=10^{52}E_{52}$ ergs is the initial fireball energy,
$\Gamma_0 = 10^2 \Gamma_2$ is the terminal coasting bulk Lorentz factor
 and $\rho=1 \rho_0$  cm$^{-3}$ is the average external
density. 

The radius of deceleration depends on the product $\rho_0 \Gamma_0^2$, so
that large variations of $\rho_0$ can be mimicked by much smaller variations
in $\Gamma$. We should remark that the ejecta Lorentz factors are
limited to $\Gamma >$ 30 (M\'esz\'aros, Laguna \& Rees 1993), and
are unlikely to exceed by much the value $\Gamma \sim 10^{3}$ \cite{err00}. 
The value of $R_{\rm dec}$ is crucial in determining whether (and at which
distance) the slower inner shells will eventually catch up with the
outer ones as they are decelerated \cite{fr00}.

For each
collision between two shells there is a reverse and a forward shock. The 
shock jump equations determine the physical parameters
of the shocked fluids, the velocity of the shock fronts and the
thickness $\Delta_{ij}$ of the merged shell at the end of the collision. We
assume that in between two consecutive collisions the thickness of the
shell increases proportionally to the fractional increase of its radius
${ d\Delta_i \over \Delta_i } \propto { dR \over R }$ \cite{SPM00}. The ejection
parameters determine the dynamics of the wind and the 
pulse dynamical efficiency, $\epsilon_{ij}$. This efficiency reflects
the differences between the Lorentz factors of a pair of colliding
shells ($\Gamma_i > \Gamma_j$). The efficiency for an
individual collision can be calculated from the initial and final bulk
energies, 
\begin{equation}
\epsilon_{ij}=1- {m_{ij}\Gamma_{ij} \over m_i\Gamma_i + m_j\Gamma_j}
\label{eff}
\end{equation}   
where $\Gamma_{ij}$ is the Lorentz factor of the resulting shell:
\begin{equation}
\Gamma_{ij}^2=\Gamma_i\Gamma_j{m_i\Gamma_i + m_j\Gamma_j \over m_i\Gamma_j + m_j\Gamma_i} ,
\end{equation}
and the resulting mass is $m_{ij}= m_i + m_j$.
 
The first collisions remove the initial random differences between the 
Lorentz factors of successive shells. If
the mean Lorentz factor $\bar{\Gamma}$ remains steady for the entire
burst duration, then the efficiency steadily decreases during the wind
expansion. If  $\bar{\Gamma}$ is modulated
on a timescale much smaller than the overall duration of the
wind, dynamically efficient collisions at large radii are still
possible. In most earlier discussions 
(see e. g. Kobayashi, Piran \& Sari 1997),
the concern was raised that internal shocks without deceleration were rather
inefficient, converting only $\la$ 25\% of 
the bulk motion energy into radiation. Since the afterglows can only
account for a few percent of the radiated energy, it was unclear
where most of the energy goes. However, Fenimore \& Ramirez-Ruiz (2000) 
showed that, for large values of $\Gamma$  (or large values of the ambient
density), deceleration is an effective catalyst for converting the bulk
motion energy into radiation.

\subsection{Turning off the central engine: a discontinuous wind}

To simulate the effects of a complete turn-off of the central site,
we impose a quiet emission time in the activity of the engine (i.e. 
the source does not emit any shells) for
about 55 seconds. Before and after this interval, 
 shells are generated at the mean rate of about 1.4 per second, with the
shell Lorentz factors $\Gamma_i$  randomly selected between
$\Gamma_{\rm min}$ ($3 \times 10^{2}$) and $\Gamma_{\rm max}$ ($9\times
10^{2}$, see Fig. 1b).

As the time of arrival of the pulses at the detector closely reflects
the activity at the central engine \cite{koba97}, it is not 
surprising that we  
find a quiescent time in the $\gamma$-ray light-curve (see
Fig. 1a), with a duration comparable to the quiet emission period at
the central engine. This is generally the case whenever the central
engine turns off for a long enough time and 
the internal shocks develop well inside the radius where the external shock decelerates
\cite{fr00}.
  
The presence of a quiet emission period in the 
central engine would divide the relativistic outflow into two well 
separated thick shells, each of them composed of many concentric inner
shells moving at different (relativistic) speeds. In section 3 we will
discuss in detail the clear signatures observable in the 
afterglow emission when the relativistic flow is discontinuous.

\subsection{Modulating a continuous relativistic wind}

There are at least two simple mechanisms which might lead to a period
of quiescent emission in the observed light curve without postulating any
quiet phase in the central engine.  The simplest possibility is that
the central engine ejects consecutive shells moving with Lorentz
factors that are essentially constant over a certain period of time,
$\Delta t_{\Gamma={\rm const}}$. In this scenario (which will be
referred to in the following as steady outflow) the requirement is that the
difference in Lorentz factors of two  
consecutive shells $\Gamma_i$, $\Gamma_j$, ($i>j$) be small enough
($\Gamma_i-\Gamma_j < \epsilon$) for the shells not to collide until
the deceleration radius (see Eq. 3). The second possibility 
may arise when the Lorentz factors of a series of consecutive shells
monotonically decrease ($\Gamma_i < \Gamma_j$) during a certain time
interval $\Delta t_{\partial\Gamma <0}$. In both cases, an observed
quiescent time interval will be  a consequence of the 
modulation of a continuous wind.

\subsubsection{Steady outflows in relativistic ejecta}

To simulate the effects of a steady outflow, we impose a period
$\Delta t_{\Gamma={\rm const}}$ of constant velocity of the ejecta, in
between two intervals during which the shells' Lorentz factors
$\Gamma_i$ are randomly selected between $\Gamma_{\rm min}$ ($10^2$)  and
$\Gamma_{\rm max}$ ($10^3$). The value of the shells' Lorentz factor 
in the steady state phase, $\Gamma_{\rm const}$, is also chosen to lie 
between $\Gamma_{\rm min}$ and 
$\Gamma_{\rm max}$. The properties of the observed light-curve  will 
depend crucially on the value of  $\Gamma_{\rm const}$. 
In particular, it is this value that essentially 
determines if a quiescence period
is present or not in the light-curve.  
We denote as $\Gamma_{\rm ave}=(\Gamma_{\rm min}+\Gamma_{\rm max})/2$
the average  Lorenz factor of the shells randomly emitted before and
after the steady  state phase.

The light-curves for three different steady
states, $\Gamma_{\rm const} > \Gamma_{\rm ave}$, $\Gamma_{\rm const} \sim 
\Gamma_{\rm ave}$ and  $\Gamma_{\rm const} < \Gamma_{\rm ave}$ are
shown in figures 2a, 2b  and 2c, respectively, for the low density case
($\rho_0=1$), and in figures  2d, 2e and 2f, respectively, for 
a very high density case ($\rho_0=10^6$).  

In the high $\Gamma_{\rm const}$ case (Fig. 2a, 2d), the interaction of the steady flow
with the early ejected material produces a low intensity component of
$\gamma$-ray emission. This underlying smooth component decreases in
intensity as the Lorentz factor of the steady flow becomes
progressively smaller (see Fig. 2b, 2e) and, depending on the background
level, could be interpreted as the signature of a
quiescent time. 
When $\Gamma_{\rm const}$ is much smaller than the average Lorentz factor of the
merged shells (Fig. 2c, 2f), 
the photon emission takes place later, producing
a much clearer quiet interval in the time history. If the central
engine ejects relativistic matter with a constant
energy rate (so that the slower shells are more heavily loaded with baryons), 
one would expect the later observed emission to be more
energetic due to the super-imposition of the two contributions (see
Fig. 2c, 2f).

Clearly the level of the underlying smooth component
associated with the  steady phase increases with increasing the value
of the product $\rho_0\Gamma^2$. Thus, for a fixed range of ejecta
Lorentz factors, the presence of quiescent times in  $\gamma$-ray
light-curves is more evident in low density
environments.   
 
The exact dynamics of these different scenarios can be more easily understood
with the help of Figure 3, which shows the radii $R_c$ 
where the collisions between shells take place, together with the 
resulting Lorentz factors $\Gamma_{ij}$, 
as functions of the time of arrival $T_{\rm toa}$ at the detector 
(see Eq. 1).

\begin{figure}
\vbox to95mm{\vfil 
\psfig{figure=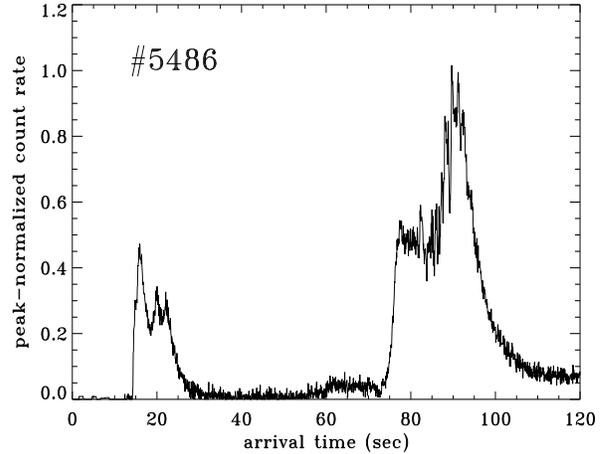,angle=0,height=70mm,width=90mm}
\caption{The time history of the long BATSE burst \#5486. Note the presence
of narrow peaks throughout the time history over an underlying
smooth emission that seems to be increasing with time. The
presence of such quiescent time may arise from a central engine which emits
relatively slow shells during its steady phase, as shown in figure 2c.}
\vfil}
\label{fig4}
\end{figure} 

At the end of
the first ejection phase of shells with random velocities, 
we are left with a rather ordered flow in which the faster (merged) shells
are the outermost ones and the slower ones follow behind.
The dynamics of the ejecta are mainly determined by the value of the 
deceleration radius (Eq. 3), which in turn depends on the external density 
and on the ejecta Lorentz factors. For a given $\Gamma_{\rm 0}$, the
denser the  external environment the smaller the radius of
deceleration,

\begin{equation}
R_{\rm dec} \sim \left\{ 
\begin{array}{ll}
 10^{16} {\rm cm} & {\rm if} \;\;\rho_0 \simeq 1 \\
 10^{14} {\rm cm} & {\rm if} \;\;\rho_0 \simeq 10^6,\\ 
\end{array} 
\right. 
\end{equation}
(see Eq. 3, where we have assumed $E_{52}=1$ and $\Gamma_2=3$).

Consequently, if the source is surrounded by a high density medium, 
a substantial fraction of the internal collisions would
take place at $R_{\rm dec}$ as the ejecta interact with the rapidly
decelerating outer shell. The merged shells produced by these
collisions will have relatively low Lorentz factors $\Gamma_{ij}$.
 
\begin{figure*}
\vbox to125mm{\vfil 
\psfig{figure=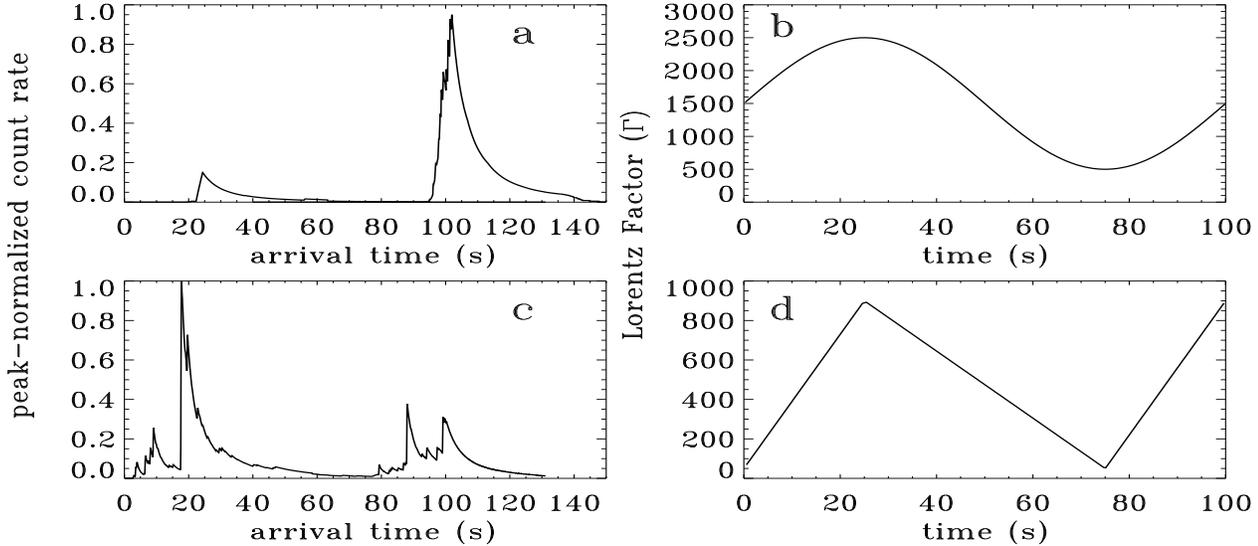,angle=0,height=105mm,width=183mm}
\caption{Simulations of what happens when the central engine modulates
the mean  
Lorentz factor in an internal shock model ($\rho_0=1$).
Panel (a) shows the light-curve due to 
a wind modulated by a sinusoid with period $P=T_{\rm dur} = 100$ s, as
shown in panel (b). Panel (c)  shows the light-curve due to 
a  wind modulated by a triangular function,
as shown in panel (d).}
\vfil}
\label{fig5}
\end{figure*}

If $\Gamma_{\rm const} > \Gamma_{\rm ave}$, the fast steady outflow 
will overtake the early emitted ejecta at a large radius (see Fig. 3a),
thermalizing their energy and boosting the Lorentz 
factor of the shell (see Fig. 3b). When the external density
is higher, the  fast 
steady outflow will overtake the decelerating shell at a radius
$R_c \ga R_{\rm dec}$ (see open squares in Fig 3a). The
resulting Lorentz factor in this case will be much lower. The
observed time over which internal shocks transform the relative
kinetic energy of the two colliding layers is of the order of
$R_c/c\Gamma_{ij}^2$. Thus,  it is the emission from these collisions at large
radii (and/or small Lorentz factors) that  
`fills the gap' in the $\gamma$-ray light-curves (see Fig. 2a and 2d)
with a number of low intensity, but large width, pulses.
 
A similar behaviour is seen when the Lorentz factor of the
steady state is close to the average value $\Gamma_{\rm ave}$.

Finally, when the
source ejects very slow shells ($\Gamma_{\rm const} < \Gamma_{\rm ave}$) 
during the steady state phase, the
expected behaviour can lead to two different scenarios.

If the burst is triggered in a low density environment, the later,
faster emission will overtake the steady ejecta before they
reach the decelerating external shock (see filled circles in Fig. 3e).
If instead the burst is triggered in a dense environment, 
the earlier ejected (faster moving) shells will begin to interact with
the surrounding medium and decelerate before
they are caught up by the later shells. Figures 3e and 3f show the
collisional parameters $R_c$ and $\Gamma_{ij}$ for this scenario in
the two cases.

In our example, the values of the ejecta Lorentz factors are such that
the later emitted shells overtake the steady
ejecta. This  causes the photon emission from these collisions to arrive later
at the detector, creating a period of quiescence in the GRB light-curve 
of duration comparable to the duration of the steady state phase of the flow
(see Fig. 2c and 2f). 
Furthermore, the late pulses arriving from these collisions give rise
to a higher underlying smooth level after the quiescence period.

Figure 4 shows the
BATSE time history of the burst 5486, that has the characteristics 
foreseen in the latter
scenario. It has many narrow peaks throughout its time history and an
underlying smooth component that seems to be increased after the 
second emission episode. However, it
is also possible that
the apparent increase of the background emission is due to a 
continuing underlying source.

\subsubsection{Winds with monotonically decreasing speed}

The second possibility  for a modulated wind to produce a quiescence period 
in the observed $\gamma$-ray light-curve, is to have in the flow a
series of shells  with  monotonically decreasing Lorentz factors.
We use two different functions to model  such modulation: 
a sine function with period $P=T_{\rm dur}$ (Fig. 5b), and
a triangular one (Fig. 5d), which accounts for a power-law decline of the 
Lorentz factors. In both cases, the time ($\Delta
t_{\partial\Gamma <0}$) during which consecutive ejected shells have
decreasing velocities is equal to ${T_{\rm dur} \over
2} = 50$ s. 

As can be seen in Fig. 5a, the sine modulation is more likely 
to give rise to a temporal
profile that shows precursor activity, while in
the case of a triangle modulation (Fig. 5b), 
the first episode of emission can be more
intense.

\subsection{X-ray emission from internal shocks}

The process by which the dissipated energy is finally radiated depends
on the energy distribution of protons and electrons in the shocked
material and on the values of the comoving density and magnetic
field. The internal shocks heat
the expanding ejecta, amplifying the preexisting magnetic field or
generating a turbulent one, and accelerate electrons, leading to
synchrotron emission and inverse Compton scattering. 
Rees \& M\'esz\'aros (1994), Papathanassiou \&  M\'esz\'aros (1996)
and Sari \& Piran (1997) calculated the radiation spectrum assuming
 that the electrons  come into
(at least partial) equipartition with the protons. If a fraction
$\varepsilon_e$ of the dissipated
energy goes into the electrons, their characteristic Lorentz factor is
given by $\Gamma_e \sim \varepsilon_e \epsilon_{\rm dis}/m_ec^2$, 
where $\epsilon_{\rm dis}$ is the 
dissipation efficiency, i.e., the amount of kinetic energy that is
converted into internal energy (a more realistic treatment, that considers
a power-law distribution for the accelerated electrons can be found in
Spada, Panaitescu \& M\'esz\'aros 2000).
The magnetic field $B$ is parameterised by the fraction
$\varepsilon_B$ of the energy of the shocked gas that it contains:
$B^2={\rm 8}\pi 
\varepsilon_B \epsilon_{\rm dis}n_e'm_pc^2$, where $n_e'$ is the comoving
frame electron number density of the shocked fluid. Assuming
$L = 10^{52}$ erg s$^{-1}$ and  $\varepsilon_B = 1/3$ (complete equipartition
between protons, electrons and the magnetic field), the value of the
magnetic field at the radius $R_c \sim c \Delta T \overline{\Gamma}^2 
\simeq 2.7 \times 10^{14} (\Delta T /1 \;{\rm
s})(\overline{\Gamma}/300)^2$ cm, where most of the
collisions take place, is $B_{\rm eq} \sim 10^3-10^4$ G
\cite{daigne98,papa96}, also
depending on the ratio $\Gamma_i / \Gamma_j$ (Eq. 5). 
Synchrotron emission will occur at a typical
energy, $h\nu_{\rm syn}$ (in the observer frame); $\gamma$-rays can be
produced by inverse Compton scattering of the synchrotron photons on the 
accelerated electrons at 
a typical energy \cite{daigne98}
\begin{equation}
h\nu_{\rm IC} \simeq h\nu_{\rm syn}\Gamma_e^2 \sim
500\left({\Gamma_{ij}  \over 300} \right)\left({B_{\rm eq} \over
1000 \, {\rm G}}\right)\left({\Gamma_e \over 100}\right)^4 {\rm keV}.
\end{equation}

The spectrum of the overall burst is the sum of all the contributions
from the individual collisions in the internal shock scenario. The
typical radiation energy of each pulse then depends on the resulting
Lorentz factor, $\Gamma_{ij}$.   
Thus, it is possible that the presence of a quiescent time in the
$\gamma$-ray light-curve is due to the shifting of the internal
shock emission into longer wavelength bands. This will happen when the
value of the resulting Lorentz factor (for a fixed luminosity) is
sufficiently small or when the luminosity drops substantially for a
certain period of time (given a constant mass loss rate). However,
when the optical thickness of the emitting shells is greater than
unity, the photons are down-scattered by the cold electrons before they
escape, leading to a further decrease in their energy. The optical
thickness in turn is determined by the wind luminosity $L$, the range
of Lorentz factors of the wind (which determines the collision radii), and the
density of the external medium. 

\begin{figure}
\vbox to100mm{\vfil 
\psfig{figure=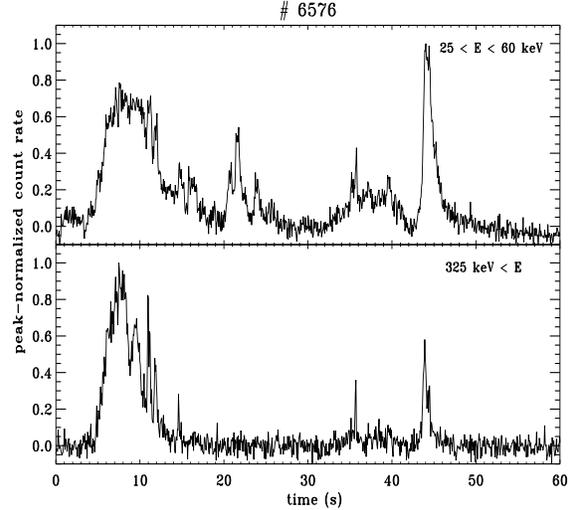,angle=0,height=70mm,width=80mm}
\caption{The time history of the long BATSE burst \#6576. Note the
presence of a quiescent interval that is near background in the
highest energy channel but that is absent in the low energy one. Thus
a measurement in a single energy channel is inadequate for the correct
identification of a complete turn-off of the central
engine. Additionally, due to the limited spectral range of BATSE a
non-detection in all four energy channels does not necessarily indicate a
complete absence of central engine activity.}
\vfil}
\label{fig6}
\end{figure}

Indeed, in the case of a very high density environment, the
underlying smooth component shown in Fig. 2 and discussed in section
2.3.1, will tend to be softer than the overall burst 
emission. Depending on the exact values of the magnetic field and the
dynamical parameters, this underlying feature may disappear from the
$\gamma$-ray emission and appear as hard X-ray radiation. For example,
from the simulation described in Figure 3c and 3d (open squares), assuming
$B \sim 10^3$ G and  $\Gamma_e=100$, the energy of the resulting
pulses responsible for the underlying smooth component (with Lorentz factors,
$\Gamma_{ij} \sim 25 - 50$), will be $h\nu_{\rm IC} \simeq 30 - 70$
keV. Figure 6 shows the BATSE time history of the burst 6576 which 
may be consistent with the above scenario, in that there is a period of quiet
emission only in the high energy channel. An inspection of the high
energy channel alone might have lead one to postulate the presence of
a quiet time in the central engine activity, but even a cursory
examination of the low energy channel reveals that this is not the
case (note that BATSE high
resolution light-curves are obtained in the 
$25 - 1000$ keV range). Alternatively, in the low density environment and
for the $\Gamma_{\rm const} < \Gamma_{\rm ave}$ case (see filled
circles in Fig. 3e and 3f), the emission episode following the quiescent time
will be much softer than the preceding one.

Observations of the prompt X-ray emission in GRBs exhibiting
quiescent times
in their light-curves can be used to constrain the inner engine 
emission properties. Any quiescent time detected simultaneously
in the X-ray and $\gamma$-ray bands will imply that, for a
comparable time, the central source has either completely switched off
or entered a steady state phase characterised by a very small
luminosity (and/or very small Lorentz factor $\Gamma_{\rm const}$).
As an example, we mention GRB 990510, detected by BeppoSAX and BATSE
\cite{kuul00}, that shows no sign of emission between 40 and 700 keV
for a period of about  30 seconds.

\subsection{Continuous vs discontinuous } 

The main conclusion we can draw from the simulations presented above
is that, in the framework of the internal shocks model, there are 
realistic assumptions that produce a long
quiescent interval in a GRB light-curve, without
having to postulate that the central source itself turns off  for a
comparably long time.

Of course, as shown if Fig. 1a, a quiet emission event is {\it always}
observed if the central engine turns off for a long enough time (provided
that little 
deceleration is occurring, see Fenimore \& Ramirez-Ruiz 2000), 
and the duration of the observed period of quiescence in the $\gamma$-ray 
light-curve would correspond to the duration of the quiescence time in
the central source.

On the other hand, for the wind modulations we have discussed, this
 is not always the case.
For the steady state modulation, as already pointed out, the actual duration
of the observed quiescent time will depend on the Lorentz factor of the steady
flow. In the most favourable case ($\Gamma_{\rm const} <
 \Gamma_{\rm ave}$, Fig. 2c), we
have simulated $10^{3}$ bursts, with a fixed
$\Delta  t_{\Gamma={ \rm const}}$, and found that the average quiescent time
 in the light-curve is shorter than $\Delta  t_{\Gamma={ \rm const}}$ by a  
factor of about 1.3. 
For the sinusoidal and triangular modulations, again 
we simulated $10^{3}$
temporal profiles for each of the two cases, 
keeping fixed $\Delta t_{\partial\Gamma <0}$. The resulting 
quiescent times in the 
$\gamma$-ray light-curve are on average shorter than  $\Delta
 t_{\partial\Gamma <0}$ by almost a factor of $2$.  
This means that, in order to 
produce an observed quiescent period of, for example, 30 seconds, 
the source has to emit a series of shells with decreasing Lorentz factors for 
about a minute. This poses severe constraints 
on any dynamical model for the inner engine.

How, in the light of
our internal shocks simulations,  is it possible to discriminate between the 
different scenarios, and in particular, to determine if the central source 
really turns off in correspondence to an observed quiescent time?

As we have seen, each of the cases described above has its own
characteristic features; however, these are not always unique.
For example, the sine modulation is more
likely to give rise to a temporal profile that shows precursor
activity (see Fig. 5a), while the `steady phase' modulation can
produce almost any observed quiescent time profile, as a consequence of
the complex dynamics of the relativistic ejecta. This is mainly due to the
large amount of free parameters that the internal shock model
offers to describe the great variety of GRBs.

However, if the central source really  turns off for a long period
when a quiescent time is observed in the $\gamma$-ray and X-ray
light-curves, the
relativistic flow will be discretised in a number of thick shells,  
each one of size roughly comparable to the duration of the 
different emission episodes 
observed at the detector. This will not be the case if such features
arise from the modulation of the relativistic ejecta, because all
these scenarios envisage the
ejection of a continuous outflow over the whole duration of the
main event at the central engine.

For this reason it is of the utmost importance to obtain as much
information as possible on the nature of the relativistic flow, as this would
provide us with some of the best clues on the stability properties of the GRB
engine (see Paper I). 
Thus, in the following sections, we study the expected
prompt (early afterglow) multi-wavelength signal. We show that this
early signal could help us determine whether
or not a gap in the
$\gamma$-ray light-curve was caused by a turn-off of the central
engine as opposed to a modulation of the 
relativistic wind.

\section{The Early Afterglow}

The internal shocks we discussed in the previous section are produced 
by the collisions of different components of the relativistic outflow 
travelling at different velocities. When the ejecta run into the
external medium two more  shocks are produced: a
short-lived reverse shock, traveling through  the ejecta; and a
long-lived forward shock propagating in the swept-up ambient
material. In the following sections we describe the expected
emission from the two external shocks (forward and reverse), focusing 
in particular on the distinctive signature to be expected from GRBs exhibiting
a quiescent time in their light-curves.

\subsection{The forward shock}

The synchrotron spectrum
from relativistic electrons that are continuously accelerated into a
power law energy distribution comprises four power-law
segments, separated by three critical frequencies: the self absorption 
frequency ($\nu_{\rm sa}$) the cooling frequency
($\nu_{\rm c}$) and the characteristic synchrotron frequency ($\nu_{\rm m}$)
\cite{spn98,mes98}. 
The spectrum and the light-curve of an afterglow are determined by 
the time evolution of these frequencies, which in turn depends on the
 hydrodynamical evolution of the fireball. 
The main temporal and spectral features of the
expected afterglows have been discussed in M\'esz\'aros, Rees \&
Wijers (1998); Sari \& Piran (1999). 

Here, we explore the hydrodynamics of the
relativistic fireball expanding in a uniform external medium, 
and the evolution of the
bolometric luminosity, without considering any spectral characteristic.  
The treatment in this section is approximate
and correction factors may need to be included in a more precise
treatment.

The interaction of the outer shell with the external medium is described by the
adiabatic Blandford-McKee (1976) self-similar solution.
In the early afterglow, as the shell progressively
collects material from the external medium, the Lorentz
factor initially stays constant. Due to the increase in the area of the shell,
 the internal energy increases with time as $\sim t^2$. 
Assuming that the
cooling is fast, the observed luminosity is proportional to the
internal energy, and so $L \sim t^2$. 

After this phase, the evolution can be of two types
\cite{sari97}, depending on the
thickness of the wind shell, on its average Lorentz factor
and on the external density; 
shells satisfying
$\Delta > (E_0 / \rho_0 m_{p} c^2)^{1/3} \Gamma_0^{2/3}$ are 
considered thick, otherwise they can be regarded as thin \cite{sprs} 
(here $E_0$
is the energy of the wind shell sweeping up the external density). 

Thick shells,
which are usually associated with long bursts, 
start a decelerating phase, with
$\Gamma (t) \sim t^{-1/4}$, after a time \cite{sari97}
\begin{equation}
t_N \sim \left({E_0 \over \Delta \rho_0 m_{p} c^4 \Gamma_0^8}\right)^{1/2},
\end{equation}
where $\Delta$ is the thickness of the wind shell.
During this phase, the luminosity is
constant and its value is about $L \sim cE_0 / 2\Delta$. This behaviour
will continue until the shell has given the surrounding material an energy comparable
to its initial energy, then a transition
to a faster deceleration phase follows,	 
with $\Gamma (t) \sim t^{-3/8}$. This second transition occurs at a time 
\begin{equation}
t_{\rm dec} \sim \left( {E_0 \over \rho_0 m_{p} c^5 \Gamma_0^8}\right)^{1/3}.
\end{equation} 
In this late phase the observed luminosity
decreases with time as $L \propto t^{-1}$. 
  
Thus, in the case of long bursts (thick wind shells), 
the signals from the internal shocks and from
the early forward shock afterglow  
overlap, since most
of the energy is extracted during a time of duration
$\Delta/c$. It is worth noting that for thin shells, 
which usually correspond to short bursts,
there is no intermediate stage of constant luminosity \cite{sari97}.
 
If the main $\gamma$-ray burst is produced by 
internal shocks, then the width of
the wind shell can be inferred directly from the observed main
burst duration: $\Delta=cT_{\rm dur}$. However, if the central engine turns
off for a certain period of time, the relativistic outflow will 
be discretised, causing
the width of the wind shell responsible for sweeping up the
external medium to be smaller.
Thus, apart from the differences in the main burst properties already
discussed in section 2 (c.f. Fig. 1 and 2), one can expect a difference in  
the relevant transition times of the forward shock evolution   
(Eqs. 8 and 9).
 
\begin{figure}
\vbox to150mm{\vfil 
\psfig{figure=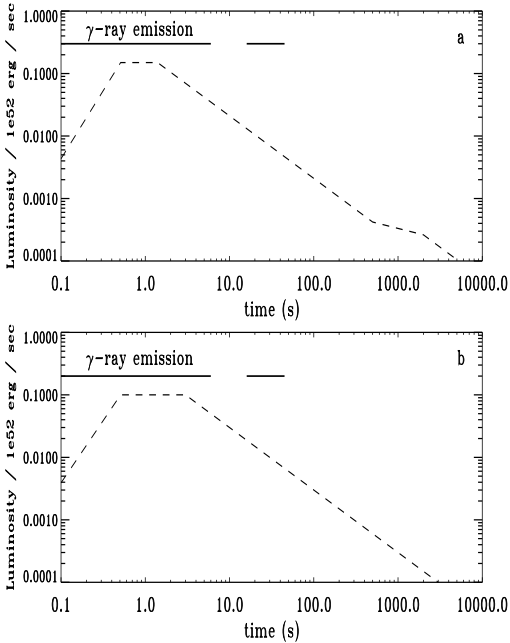,angle=0,height=95mm,width=80mm}
\caption{Bolometric luminosity of the shocked ISM ($\rho_0=1$; dashed line) 
as a function of time for
two different mechanisms of gap production in a $\gamma$-ray light-curve
(solid line). The case of a central engine that
turns off is shown in the upper panel (a), 
while the case of a modulated continuous wind is shown in the lower one (b). 
At early stages the Lorentz factor is constant, and the luminosity
increases due to the increase of shell area. When the shell 
has given the ISM an energy comparable to its initial energy, a
self-similar solution is established. In this phase the luminosity
drops as 
$t^{-1}$. An intermediate state, $t^{-1/4}$, may occur for thick
shells, leading to constant luminosity. Both
cases use $E_{\rm tot}=10^{53}$ ergs, $\Gamma_{0} = 500$ and the internal
Lorentz factor within the flow are chosen as in Fig. 1 (for the case (b)
$\Gamma_{\rm const} = 200$). }
\vfil}
\label{fig7}
\end{figure}

Figure 7 illustrates the two
different afterglow behaviours we expect for a GRB exhibiting a
quiescent time of ten seconds between $t=6$ s and $t=16$ s.
We have fixed $\Gamma_{0} = 500$ and the internal Lorentz 
factors within the flow are chosen as in Figure 1. The external density
 has been chosen to be the typical ISM one ($\rho_0=1$). 
In the first case (Fig. 7a), the observed period of quiescence is 
produced by a central engine that
turns off (as discussed in section 2.2); in this scenario,  
the relativistic outflow is divided into two well separated thick
shells, an outer and an inner one,
 of thickness $\Delta_{\rm out} = 1.8 \times 10^{11}$ cm and 
$\Delta_{\rm in} = 8.7 \times 10^{11}$ cm, respectively. In the
alternative scenario (Fig. 7b), the observed quiescent time is due to
a  steady state modulation of a continuous  relativistic 
wind (as discussed in section
2.3.1), composed of a single shell of thickness  $\Delta = cT_{\rm dur} = 1.35
\times 10^{12}$ cm.  
In both cases we have fixed $E_{\rm tot} = 10^{53}$
ergs, but this energy is shared differently by the different shells
sweeping up the external medium: 
in the continuous wind case we have $E_0=E_{\rm tot}$, while
in the discontinuous one  the energy of the outer shell is 
$E_{0, {\rm out}} = E_{\rm tot} \Delta_{\rm out}/(\Delta_{\rm out}+\Delta_{\rm in}) = 0.17 E_{\rm tot}$.  
The difference in the thickness of the outer
shell (the one that interacts with the surrounding medium) is
responsible for  the dissimilarity in duration of the constant luminosity
phase of the afterglow light-curve.
 
When a quiescent time in the main $\gamma$-ray emission is observed 
after a very short pulse, the difference $t_{\rm dec} - t_{N}$ 
in the two cases may be large enough
to allow us to establish whether the central engine has turned off or
not by 
looking at the detail of the long wavelength afterglow emission. 
However, if the quiescent time appears after a long period of
emission, such discrimination may be very difficult.

We would also like to emphasise that if the burst is located in 
a denser environment ($\rho_0 \ga 10^4$), 
expected if the source is associated with the collapse of a massive star,
the hydrodynamical evolution would take place on smaller timescales
(see Eqs. 8 and 9) and the bolometric luminosity will start declining
after a few hundredths of a second.

Furthermore, in the case in which the central engine turns off 
to produce the observed quiescent time,
one expects a collision between the outer shell (which, after the internal 
shocks have taken place is left with an energy
$E_{\rm out}$) and the inner one (with energy $E_{\rm in}$). The overall
effect of the collision will be, at a fixed frequency, the increase of
the flux by a factor of $ \sim (1 + E_{\rm in}/E_{\rm out})^{1.4}$ 
\cite{kp00} after about 10$^{3}$ seconds, as
shown in Figure 7a. Once more, if the external medium is much denser, this 
bump  in the light-curve would be observed earlier, as the collision will take
place closer in, its exact location depending on the external density
and on the dynamical  parameters of the two shells \cite{kp00}.

 From Figure 7 it is also clear that the signals from the
internal shocks (the main GRB, see Fig. 8 for the light-curve) and the
early forward shock emission overlap. Therefore, it might be difficult to
detect the smooth external shock component and probably even more difficult
to discriminate between the two scenarios by looking 
for any increasing underlying background component in the $\gamma$-ray 
light-curve (as discussed in section 2.3.1 and in Fig. 4).

In the next section we will show that the early detection of the
prompt  emission from the reverse shock could in principle be a much
clearer test that would help us to discriminate between the two alternative
scenarios.

\subsection{The reverse shock}

The reverse shock gives the right magnitude for the observed 
prompt optical flash with
reasonable energy requirements of no more than a few $10^{53}$ ergs
emitted isotropically \cite{mes97}. The ejecta cool adiabatically
after the reverse shock has passed through and settle down into a
part of the Blandford-McKee solution that determines the late profile
of the shell and the external medium. Thus, unlike the continuous 
forward shock emission, 
the reverse shock terminates once the shock
has crossed the shell and the cooling frequency has dropped below the observed
range. The reverse shock contains, at the time it crosses the
shell, an amount of energy comparable to that in the 
forward one. However,
its effective temperature is significantly lower (typically by a
factor of $\Gamma$). Using the shock jump
condition and assuming the electrons and the magnetic field acquire a
fraction $\varepsilon_e$ and $\varepsilon_B$, respectively, of the
equipartition energy, one can
describe the hydrodynamic and magnetic conditions behind the shock.
For the reverse shock, the two frequencies that determine the spectrum, 
the cooling  
($\nu_{\rm c}$) and the synchrotron ($\nu_{\rm m}$) one,
are easily calculated by comparing them to
those of the forward shock \cite{mes97,mr99,sprs}. Assuming that the
forward shock and the reverse shock move with a similar Lorentz
factor, the reverse shock frequency at the peak time, $t_p$, is given by
\begin{equation}
\nu_p \simeq 2.1 \times 10^{14} \left({\varepsilon_e \Gamma_0 \over 30}\right)^2
\left({\varepsilon_B \over 0.3}\right)^{1/2} \rho_0^{1/2} {\rm Hz},
\end{equation}
which, in the low density case, favours strong optical emission
\cite{sprs}. The peak time can be estimated with the help of 
$t_p= {\rm max}[ t_{\rm dec}, \Delta /c]$. 
After this time  self-similar evolution begins. Under
the flux-freezing field behaviour, for an adiabatic case ($\Gamma
\propto r^{-3/2}$) and an electron index $p=2$, the photon spectral
index above $\nu_{\rm m}$ is $\beta = - 1/2$ ($F_\nu \propto \nu^\beta$).
The spectral flux, $F_\nu$, has an approximate time dependence
of $t^{-2.1}$ (see M\'esz\'aros \& Rees 1999 for different
spectral behaviours), which is in rough agreement with the 
ROTSE observations of GRB990123 
(Akerlof et al. 1999 report a $t^{-2}$ dependence for
about 600 seconds). 

Assuming this dependency, in Figure 8 we show the
reverse shock afterglow emission superimposed on the $\gamma$-ray light-curve
for the two different scenarios of
gap production described in the above section. 
In Figure 8a we show the emission from a turned-off central source, 
while in Figure 8b we show the emission from  a continuous modulated 
relativistic wind, both giving rise to a quiescent time of 
comparable duration in the $\gamma$-ray time history.
 
Clearly, since the
width of the shell responsible for sweeping up the
external medium varies by a factor of $\Delta / \Delta_{\rm out} = 7.5
$ from one scenario to the other, one expects 
the reverse shock emission to peak 30 seconds earlier if
the central engine turns off after six seconds. 
Furthermore, a second peak in the reverse shock
emission may appear when the inner shell collide with the outer
shell. This inner shell would then crash into the reverse shock,
thermalizing its energy and boosting the power of the prompt
afterglow. This second feature 
can be observed if the collision occurs before the reverse shock
crosses the shell. 
It is worth noting that, while the prompt afterglow emission will
be blueshifted in the EUV/soft X-ray bands if $\rho_0 \ga 10^4$ (see Eq. 10),
the peak time $t_p$ will not be strongly affected by the value 
of the density of the external medium ($\Delta/c \gg t_{{\rm dec}}$, in this
case).

\begin{figure}
\vbox to120mm{\vfil 
\psfig{figure=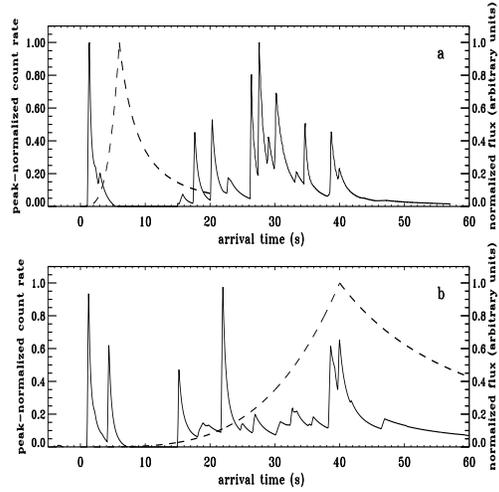,angle=0,height=80mm,width=80mm}
\caption{Light curves in two different energy bands for the two
different scenarios described in figure 7. The solid curve is the main burst
produced by internal shocks. The dashed line is the
emission from the reverse shock. This emission terminates once the
reverse shock crosses the shell and the cooling frequency drops below the
observed frequency. During this period of emission one expects a $
\sim t^{-2}$ dependence. For a central engine that
turns off (panel a), the emission from the reverse shock peaks much
earlier than in the case of an engine in a steady state (panel b).} 
\vfil}
\label{fig8}
\end{figure}

We can thus conclude that the multiwavelength signature from
the reverse shock emission, if measured,
would be a clear way of determining whether or not the central engine
turns off in order to produce a quiet period in the $\gamma$-ray
time history.

\section{Observed Properties of Quiescent Times}

In a previous work (Paper I) we have studied possible correlations 
between the duration of a quiescent time and that of the adjacent 
emission episodes. We have shown that a strong quantitative 
proportionality relation exists between the
duration of an emission episode and the quiescent time elapsed since
the previous episode; while we found no clear correlation between the
length of an emission
episode and that of the following quiescent time.

In Paper I we outlined the general properties any dynamical 
system has to possess for it to show the observed correlation, under
the hypothesis that the central engine turns off 
in correspondence to a quiescent time. Moreover,
we envisaged that the mechanism responsible for extracting and
dissipating the energy has to take 
place in a meta-stable configuration, such that the longer the
accumulation period, the higher is the stored energy available for the
next episode.

The hypothesis of an intermittent central engine, although intriguing, 
has to be tested against the observations. As we have demonstrated here,
in the internal-external shock model, a modulation of the relativistic flow 
can  produce quiescent times in the observed $\gamma$-ray
lightcurve. Furthermore, a slow steady outflow ($\Gamma_{\rm
const}<\Gamma_{\rm ave}$),  
preceded and followed by a much faster, randomly modulated wind, could
possibly be the origin of the observed correlation.
As can be deduced from Figure 3d, if the time over which the source emits 
steadily is increased, the number of collisions  with the later 
emitted shells (and the energy extracted from 
these collisions) will be larger. The whole energy contained in the
steady outflow will be detected simultaneously or after the
emission produced by the internal collisions between the following
faster emitted shells.  

It is thus reasonable to 
assume that the emission episode that follows a quiescent time produced 
in this manner will be more energetic.
Nonetheless, the tight correlation we observe would require additional
tuning of the parameters, as the  actual duration of the emission 
episode that follows a quiescent time is 
determined by the original duration of the emission at the
inner engine and by the relative value of the Lorentz factor of the steady
outflow.   

In order to extract as much information as possible 
from the correlation, 
and to gain more insight into the dynamics 
of GRB central sources, 
we first need to better understand the physical processes responsible for 
the production of quiescent times in GRB light-curves. The fundamental 
question we would like to answer is whether the gaps are produced by a
turning-off of the central engine or by a 
structure in the relativistic outflow
velocity space.

In the former case, while it would be 
easy to account for the observed episodes of quiescence,
we would need a model of the central engine able to go dormant 
for a period which is 
long compared to the  typical dynamical timescale
($\sim$ milliseconds) of the central source, and to be active again afterwards.
Then, the observed correlation could be a 
clue to unveil the dynamical (and stability) properties of the system, 
as discussed in Paper I.
 
Alternatively, if the quiescent times are due to modulations of 
a relativistic wind that produces the $\gamma$-ray 
emission via internal shocks, 
they will instead probe the velocity structure of the outflow, and 
possibly its interaction with the ambient medium.

\section{Conclusions}

The very existence of quiescent times in GRB light-curves poses severe
restrictions on the emission models. It is well known, for example,  
that in the
external shock scenario it is almost impossible to reproduce this
property \cite{fr99}. 
In the internal shocks scenario the actual temporal profile is the
outcome of the complex dynamics of the relativistic outflow, which has usually 
been considered as sequence of shells moving at different speeds.
Within this framework, it is relatively easy to accommodate a large variety of 
GRB temporal profiles, given the large number of parameters on which the 
final observed light-curve depends.  

The work we have presented here has been done with the aim of 
finding possible observational tests that could help us to discriminate 
between a turning-off of the central engine or a continuous
relativistic outflow.
We studied the various 
mechanisms that can produce a quiescent time in the internal shock model, 
and we have presented and discussed 
the results of a number of burst simulations.  
We conclude that, within such models, a central engine 
that goes dormant for a long 
enough period will always produce a quiescent 
time in the $\gamma$-ray light-curve
of comparable duration. However,  we have also shown that 
internal shocks can produce
significant long periods of quiescence in the GRB light-curves
without the central engine having to switch off.
This can be achieved by an opportune modulation of a continuous 
relativistic wind. We discuss how 
different modulations could in principle be distinguished by studying the
properties of burst light-curves in the X-ray and $\gamma$-ray energy
bands (see Figs 4 and 6). 

It is moreover possible to observationally 
determine whether  the central engine turns off or not 
by analysing the  multi-wavelength afterglow emission.
In particular, we have shown that the peak of the prompt
afterglow emission from the reverse shock strongly depends on whether
the relativistic  ejecta are part of a continuous wind or are instead
made of a number of discrete  thick shells, each one corresponding to
an emission episode in the burst time history.   
To this end, a very rapid optical follow-up of the long bursts 
exhibiting periods of quiescence would be of the utmost importance, and could 
improve our understanding of the dynamical properties of $\gamma$-ray bursts
progenitors.

\section*{Acknowledgements}

We thank P. Natarajan, A. Celotti, G. Morris and P. Madau for useful
comments and suggestions. We are particularly grateful to
E. E. Fenimore and P. M\'esz\'aros for very helpful insight regarding
internal-shock calculations.  
ERR acknowledges support from
CONACYT, SEP and the ORS foundation. AM thanks PPARC and the TMR
network `Accretion onto black holes, compact stars and protostars',
funded by the European Commission under contract number
ERBFMRX-CT98-0195, for support. MJR acknowledges support from the
Royal Society.

\bsp

\label{lastpage}


\begin{thebibliography}{99}

\bibitem[\protect\citename{Akerlof et al. }1999]{rotse99}
Akerlof, C. W., et al. 1999, Nature, 398, 400.

\bibitem[\protect\citename{Blandford \& McKee}1976]{BK76}
Blandford, R. D., \& McKee, C. F., 1976, Phys. Fluids, 19, 1130.

\bibitem[\protect\citename{Costa et al. }1997]{costa97}
Costa, E., et al. 1997, Nature, 387, 783.

\bibitem[\protect\citename{Daigne \& Mochkovich } 1998]{daigne98}
Daigne, F., \& Mochkovitch, R., 1998, MNRAS, 296, 275.

\bibitem[\protect\citename{Fenimore et al. }1999]{fen99}
Fenimore, E. E., Cooper, C., Ramirez-Ruiz, E., Sumner, M. C., Yoshida,
A. \& Namiki, M., 1999, ApJ, 512, 683.

\bibitem[\protect\citename{Fenimore \& Ramirez-Ruiz }1999]{fr99}
Fenimore, E. E. \& Ramirez-Ruiz, E., 1999, PASP Conf. Proc. Gamma-Ray
Bursts: The First Three Minutes, astro-ph/9906125

\bibitem[\protect\citename{Fenimore \& Ramirez-Ruiz }2000]{fr00}
Fenimore, E. E. \& Ramirez-Ruiz, E., 2000, to appear in ApJ,
astro-ph/9909299.

\bibitem[\protect\citename{Kobayashi, Piran \& Sari }1997]{koba97}
Kobayashi, S., Piran, T., \& Sari, R., 1997, ApJ, 490, 92.

\bibitem[\protect\citename{Kouveliotou et al. }1993]{kou93}
Kouveliotou, C., et al., 1993, ApJ, 413, L101.

\bibitem[\protect\citename{Kulkarni et al. }1999]{kul99}
Kulkarni, S. R., et al., 1999, Nature, 398, 389.

\bibitem[\protect\citename{Kumar \& Piran }2000]{kp00}
Kumar, P. \& Piran, T., 2000, ApJ, 532, 286.

\bibitem[\protect\citename{Kuulkers et al. }2000]{kuul00}
Kuulkers et al., 2000, ApJ, 538,638.

\bibitem[\protect\citename{Lattimer \& Schramm }1976]{lat76}
Lattimer, J. M., Schramm, D. N., 1976, ApJ, 210, 549.

\bibitem[\protect\citename{McFayden, Woosley \& Heger }1999]{mc99}
MacFadyen, A. I., Woosley, S. E. \& Heger, A., 1999, to appear in ApJ, astro-ph/9910034.

\bibitem[\protect\citename{M\'esz\'aros, Laguna \& Rees }1993]{mlr93}
M\'esz\'aros, P., Laguna, P. \& Rees, M. J. , 1993, ApJ, 415, 181.

\bibitem[\protect\citename{M\'esz\'aros \& Rees }1997]{mes97}
M\'esz\'aros, P. \& Rees, M. J. , 1997, ApJ, 476, 232.

\bibitem[\protect\citename{M\'esz\'aros, Rees \& Wijers }1998]{mes98}
M\'esz\'aros, P., Rees, M. J., Wijers, R., 1998, ApJ, 499, 301.

\bibitem[\protect\citename{M\'esz\'aros \& Rees }1999]{mr99}
M\'esz\'aros, P. \& Rees, M. J. , 1999, MNRAS, 306, L39.

\bibitem[\protect\citename{Norris et al. }1996]{norris96}
Norris, J. P. et al., 1996, ApJ, 459, 2393.

\bibitem[\protect\citename{Paczy\'nski }1998]{pac98}
Paczy\'nski, B., 1998, ApJ, 494, L45.

\bibitem[\protect\citename{Papathanassiou \& M\'esz\'aros }1996]{papa96}
Papathanassiou, H. \& M\'esz\'aros, P., 1996, ApJ, 471, L91.

\bibitem[\protect\citename{Ramirez-Ruiz \& Fenimore }1999]{err99}
Ramirez-Ruiz, E. \& Fenimore, E. E., 1999, A\&AS, 138, 521.

\bibitem[\protect\citename{Ramirez-Ruiz \& Fenimore }2000]{err00}
Ramirez-Ruiz, E. \& Fenimore, E. E., 2000, ApJ, 539, 712.

\bibitem[\protect\citename{Paper I }2000]{rrm2000}
Ramirez-Ruiz, E. \& Merloni, A., 2001, MNRAS, 320, L25 (Paper I).

\bibitem[\protect\citename{Ramirez et al. }2000]{winds2001}
Ramirez-Ruiz, E., Dray, L. M., Madau, P. \& Tout, C., 2001, submitted
to MNRAS, astro-ph/0012396.
 
\bibitem[\protect\citename{Rees \& M\'esz\'aros }1994]{Rees94}
Rees, M. J. \& M\'esz\'aros, P., 1994, ApJ, 430, L93.

\bibitem[\protect\citename{Rees }1999]{Rees99}
Rees, M. J., 1999, A\&AS, 138, 491.

\bibitem[\protect\citename{Sari \& Piran }1997]{piran97}
Sari, R. \& Piran, T., 1997,  ApJ, 485, 270.

\bibitem[\protect\citename{Sari }1997]{sari97}
Sari, R., 1997, ApJ, 489, L38.

\bibitem[\protect\citename{Sari, Piran \& Narayan }1998]{spn98}
Sari, R., Piran, T., \& Narajan, R., 1998, ApJ, 497, L17.
 
\bibitem[\protect\citename{Sari \& Piran }1999]{sprs}
Sari, R. \& Piran, T.,1999,  ApJ, 517, L109

\bibitem[\protect\citename{Spada, Panaitescu \& M\'esz\'aros }2000]{SPM00}
Spada, M., Panaitescu, A. \&  M\'esz\'aros, P., 2000,
ApJ, 537, 824. 

\bibitem[van Paradijs et al. 1997]{vp97}
van Paradijs et al, Nature, 386, 686, 1997.

\end{thebibliography}
\end{document}